\documentclass{epl}

\title{Chemical potential shift in \chem{La_{1-x}Sr_xMnO_3}:\\ Photoemission test of the phase separation scenario}
\shorttitle{Chemical potential shift in \chem{La_{1-x}Sr_xMnO_3}}

\author{J.~Matsuno\inst{1} \and A.~Fujimori\inst{1,2} \and Y.~Takeda\inst{3} \and M.~Takano\inst{4}}
\shortauthor{J. Matsuno \etal}
\institute{
\inst{1}Department of Physics, University of Tokyo - Bunkyo-ku, Tokyo 113-0033, Japan\\
\inst{2}Department of Complexity Science and Engineering, University of Tokyo - Bunkyo-ku, Tokyo 113-0033, Japan\\
\inst{3}Department of Chemistry, Mie University - Tsu 514-8507, Japan\\
\inst{4}Institute of Chemical Research, Kyoto University - Uji, Kyoto 611-0011, Japan\\
}
\pacs{71.27.+a}{Strongly correlated electron systems; heavy fermions}
\pacs{71.30.+h}{Metal-insulator transitions and other electronic transitions}
\pacs{79.60.-i}{Photoemission and photoelectron spectra}

\begin{document}

\maketitle

\begin{abstract} 
  We have studied the chemical potential shift in
  La$_{1-x}$Sr$_x$MnO$_3$ as a function of doped hole concentration by
  core-level x-ray photoemission. The shift is monotonous, which means
  that there is no electronic phase separation on a macroscopic scale,
  whereas it is consistent with the nano-meter scale cluster formation
  induced by chemical disorder. Comparison of the observed shift with
  the shift deduced from the electronic specific heat indicates that
  hole doping in La$_{1-x}$Sr$_x$MnO$_3$ is well described by the
  rigid-band picture. In particular no mass enhancement toward the
  metal-insulator boundary was implied by the chemical potential
  shift, consistent with the electronic specific heat data.
\end{abstract}

It is generally recognized that the manganites $R_{1-x}A_x$MnO$_3$
with the perovskite structure, where $R$ is a rare-earth ($R=$ La, Pr,
Nd) and $A$ an alkaline-earth metal ($A=$ Sr, Ba, Ca), exhibit a very
rich phase diagram because of competition between the ordering and
fluctuations of the spin, lattice and charge degrees of
freedom~\cite{Mn_review}. They are antiferromagnetic (AFM) insulators
for undoped materials ($x=0$) while hole doping produces a
ferromagnetic (FM) metallic phase, which shows colossal
magnetoresistance (CMR). Historically, the ferromagnetism in this
phase has been explained by double exchange mechanism~\cite{DE}.
Materials with $x$ close to 0.5 have been the focus of recent studies
because most of the manganites show so-called $CE$-type
antiferromagnetic charge- and orbital-ordered insulating state at
$x=0.5$.  When this spin-charge-orbital ordering completes with
metallic behavior, they also exhibit a remarkable CMR effect, which
becomes the subject of intensive studies. The origin of the CMR
behaviors has been debated for many years since it was pointed out
that double exchange interaction alone is insufficient to explain the
large resistivity change~\cite{Millis}.

Recently, the possibility of phase separation as the origin of the CMR
is discussed by theoretical studies. Moreo \textit{et
  al.}~\cite{Moreo} has pointed out the possibility of the coexistence
of the antiferromagnetic or ferromagnetic insulating phase and the
ferromagnetic metallic phase and argued that this phase separation may
be responsible for the CMR effect based on their quantum Monte Carlo
simulation on the double exchange model including orbital degeneracy
and electron-phonon coupling. The tiny Drude weight in the optical
conductivity observed in the metallic region of
$R_{1-x}A_x$MnO$_3$~\cite{Mn_review} has been explained in the
framework of phase separation~\cite{Nagaosa}. The phase separation
picture has also been suggested by some experimental results: electron
microscopy studies~\cite{Uehara} have shown a phase separation on a
large length scale of $\sim100$ nm for
La$_{0.625-y}$Pr$_y$Ca$_{0.375}$MnO$_3$ ($y=0.4$), which shows
ferromagnetic metallic behavior below the Curie temperature. NMR study
of La$_{1-x}$Ca$_{x}$MnO$_3$~\cite{Papavassiliou} has revealed that
the system is separated into a ferromagnetic metallic state and a
ferromagnetic insulating state for $0.20 \leq x \leq 0.33$.

Among the manganites, La$_{1-x}$Sr$_{x}$MnO$_3$ is expected to have
the widest band width. Magnetic and transport
measurements~\cite{Urushibara} have revealed that the system becomes a
ferromagnetic metal in a wide composition range of $0.175\le x\le0.6$.
In the end member LaMnO$_3$, the $A$-type antiferromagnetism is
accompanied by a Jahn-Teller distortion of the Mn$^{3+}$ ion with
$3d^4$ configuration and the orthorhombic distortion. No clear sign of
charge-ordered state has been found for $x \sim 0.5$ probably because
of the wide band width of La$_{1-x}$Sr$_{x}$MnO$_3$. Very recently,
however, materials with $x\sim 0.5$ have been found to show different
types of ground states depending on the crystalline states. Single
crystals~\cite{Akimoto} become FM metals for $x=0.5$ and AFM metals
for $x\geq0.54$. A study using polycrystals has suggested a signature
of charge ordering for $x\sim0.5$ from the measurements of the sound
velocity~\cite{Fujishiro}. Another polycrystal study has suggested
that FM and AFM domains coexist with an anomaly due to charge
ordering~\cite{Patil}. These results suggest that charge ordering or
at least its fluctuations are important also in
La$_{1-x}$Sr$_{x}$MnO$_3$. The pair-distribution function analysis of
pulsed neutron diffraction data of La$_{1-x}$Sr$_{x}$MnO$_3$ has shown
that local Jahn-Teller distortions are present in the metallic state
up to at least 35\% of hole doping, in disagreement with the
homogeneous picture of the double exchange model~\cite{Louca}.

In general, electronic phase separation results in the pinning of the
chemical potential $\mu$. The numerical studies~\cite{Moreo,Yunoki}
have indeed shown the pinning of $\Delta\mu$ in the composition range
where the phase separation is realized. Even if there is no phase
separation on a macroscopic ($>$ 100 nm) scale, $\Delta\mu$ would be
pinned when there exists a microscopic ``phase separation'', where
hole-rich and hole-poor regions coexist on a nanometer scale and the
volume of the hole rich region changes proportionally to the doped
hole concentration, as has been demonstrated for the ``charge
stripes'' in the high-$T_c$ cuprates~\cite{Ino}. Hence, experimentally
obtained $\Delta\mu$ can be directly used to test the above
theoretical predictions. Core-level photoemission spectroscopy is a
suitable method for this purpose: the chemical potential shifts can be
deduced from the core-level shifts since in photoemission experiments
the binding energy of each core level is measured relative to the
chemical potential $\mu$. We note that one can distinguish an
electronic phase separation and a chemical inhomogeneity by this
method because no pinning of $\Delta\mu$ is expected in the latter
case. In this paper, we present the results of highly precise x-ray
core-level photoemission measurements of polycrystalline
La$_{1-x}$Sr$_{x}$MnO$_3$ ($0\leq x\leq0.6$) to study the chemical
potential shift $\Delta\mu$.

Polycrystalline samples of La$_{1-x}$Sr$_{x}$MnO$_3$ were prepared by
solid-state reaction. The sample preparation procedure is described in
detail elsewhere~\cite{Saitoh1}. X-ray photoemission spectroscopy
(XPS) measurements were carried out using a Mg x-ray source ($h\nu
=$1253.6 eV).  All the measurements were done at liquid-nitrogen
temperature except for $x=0$. Because charging effect was observed for
$x=0$ at liquid-nitrogen temperature, it was measured at room
temperature. For an accurate determination of the core-level energy
shifts, we have corrected for the time-dependent fluctuations of the
retarding voltage and the pass energy of the spectrometer by
simultaneous measurements of the voltage of the outer hemisphere, the
inner hemisphere and the entrance slit.  After the correction, the
errors in the measured shifts were reduced to $\sim10$ meV if the
measurements were made within a day. In order to achieve this accuracy
for the whole set of the samples, we performed measurements of samples
of interest and a reference sample within one day. Relative core-level
energy shifts were thus measured with referenced to $x=0.5$. The
samples were repeatedly scraped \textit{in situ} with a diamond file.
We have adopted spectra taken within 50 minutes after scraping and the
reproducibility of the spectra was confirmed by repeated scraping. We
note that even with the energy resolution of XPS ($\sim$0.9 eV), we
could determine the core-level shifts with an accuracy of $\sim$ 10
meV if the spectral shape does not change with the composition as
confirmed by the measurements of the gold $4f_{7/2}$ core-level
spectrum. In the case of La$_{1-x}$Sr$_{x}$MnO$_3$, we have achieved
the accuracy of 50--100 meV due to the spectral changes with
$x$~\cite{Ino}.

\begin{figure}[t]
  \onefigure[scale=0.4]{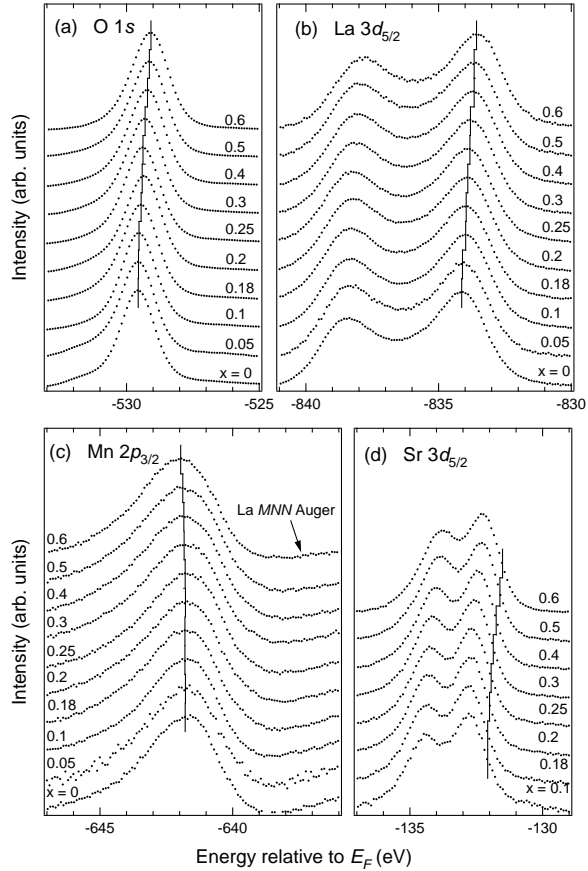}
  \caption{XPS spectra of the (a) O 1$s$, (b) La 3$d$, (c) Mn 2$p$ and
    (d) Sr 3$d$ core levels of La$_{1-x}$Sr$_{x}$MnO$_3$. Energies are
    referenced to the chemical potential $\mu$, i.e., the Fermi level
    $E_F$.}
  \label{fig:alldata}
\end{figure}

Figure~\ref{fig:alldata} shows the XPS spectra of the O 1$s$, La 3$d$,
Mn 2$p$ and Sr 3$d$ core levels. Energy shifts of those core levels
were determined by comparing the positions of the spectral features
whose line shapes do not change with composition as indicated by
vertical lines in Fig.~\ref{fig:alldata}. The shift $\Delta E$ of a
core level with varying chemical composition, measured relative to
$\mu$ is given by
\[\Delta E=\Delta\mu+K\Delta Q+\Delta V_M-\Delta E_R, \]
where $\Delta\mu$ is the change in the chemical potential, $\Delta Q$
is the change in the number of valence electrons on the atom
considered, $\Delta V_M$ is the change in the Madelung potential, and
$\Delta E_R$ is the change in the extra-atomic relaxation
energy~\cite{Ino,Hufner}. Here, $\Delta Q$ produces changes in the
electrostatic potential at the core-hole site as well as in the
intraatomic relaxation energy of the core-hole final state. $\Delta
E_R$ is due to changes in the screening of the core hole potential by
metallic conduction electrons and polarizable surrounding ions.

\begin{figure}[htbp]
  \onefigure[scale=0.6]{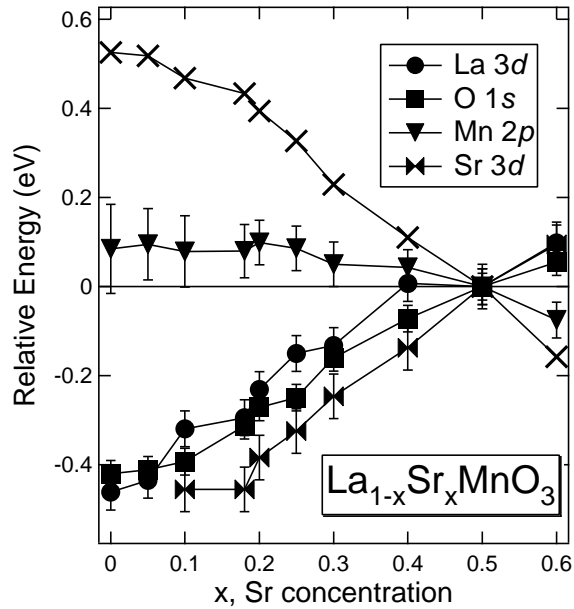}
  \caption{Energy shifts of the O 1$s$, La 3$d$, Mn 2$p$ and Sr 3$d$
    core levels relative to $x=0.5$ plotted as functions of hole
    concentration $x$. Crosses denote the difference between the shift
    of the Mn 2$p$ level and the average shift of the O 1$s$, La 3$d$
    and Sr 3$d$ levels.}
  \label{fig:allshift}
\end{figure}

One can see from Fig.~\ref{fig:alldata} that the observed energy
shifts with $x$ are common to the O 1$s$, La 3$d$ and Sr 3$d$ core
levels, whereas the shift of the Mn 2$p$ core level is opposite to
them. The line shapes of all the core levels do not change
significantly for these compositions, however, the contribution of La
$MNN$ Auger emission is not negligible in the energy range of the Mn
2$p$spectra and its intensity changes with La concentration. We have
estimated the shifts by the center of gravity for the Mn 2$p$ core
level, though the shift estimated from the peak positions give nearly
the same result. The estimated energy shifts are shown in
Fig.~\ref{fig:allshift}. Crosses denote the difference
between the shift of the Mn 2$p$ level and the average shift of the O
1$s$, La 3$d$ and Sr 3$d$ levels, demonstrating that the Mn 2$p$ shift
can be decomposed into two simple components, namely, the shift common
to all the four levels and the shift which increases the binding
energy linearly in $x$ as in the case of
La$_{2-x}$Sr$_{x}$CuO$_4$~\cite{Ino}. The extra component of the Mn
2$p$ core-level shift would be attributed to the increase in the Mn
valence with hole doping ($\propto -\Delta Q$) from Mn$^{3+}$ towards
Mn$^{4+}$, i.e., to the so-called chemical shift.  Then the shift
common to all the four levels should reflect the chemical potential
shift $\Delta\mu$ as in the case of
La$_{2-x}$Sr$_{x}$CuO$_4$~\cite{Ino}. In particular, one can exclude
the effect of changes in the Madelung potential caused by the La$^{3+}
\rightarrow$ Sr$^{2+}$ substitution because the changes in the
Madelung potential would cause shifts of the core levels of the
O$^{2-}$ anion and the La$^{3+}$ (Sr$^{2+}$) cation in different
directions.

\begin{figure}[htbp]
  \onefigure[scale=0.6]{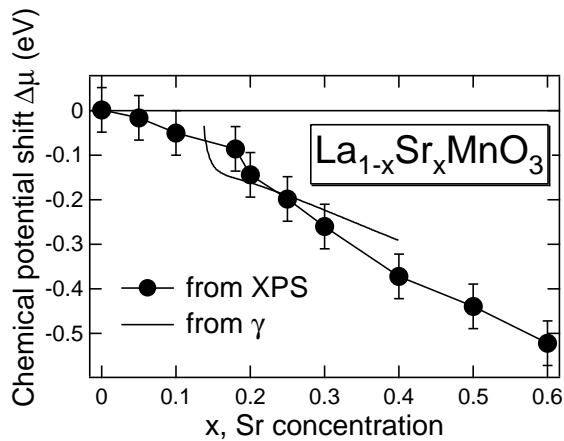}
  \caption{Chemical potential shift $\Delta\mu$ in
    La$_{1-x}$Sr$_x$MnO$_3$ as a function of Sr concentration $x$.
    The dotted line is calculated from the specific heat~\cite{Okuda}
    using Eq.~(\ref{eq:1}) with $F=0$. For details, see the text.}
  \label{fig:specific}
\end{figure}

Thus we have deduced the chemical potential shift $\Delta\mu$ from the
average of the O 1$s$, La 3$d$ and Sr 3$d$ core-level shifts as a
function of Sr concentration $x$ as shown in Fig.~\ref{fig:specific}.
For $x=0$ and $x=0.05$, we adopted the average of the O 1$s$ and La
3$d$ core-level shifts since the Sr 3$d$ core levels were not
available or too weak to determine the shift. The most important
feature in Fig.~\ref{fig:specific} is that the chemical potential
monotonously decreases with hole doping ($x$); we do not find any sign
of pinning of the shift over the whole composition range. This means
that there is no obvious electronic phase separation proposed both
experimentally and theoretically. The absence of electronic phase
separation into different spatial regions with different hole
concentrations over macroscopic length scale is quite reasonable
because it would cost huge Coulomb energy~\cite{Moreo}. Then the
``phase separation'' observed by electron microscopy~\cite{Uehara}
would be most likely due to chemical inhomogeneity. Also, electronic
phase separation on a microscopic scale as in the stripe phase of the
high-$T_C$ cuprates can be excluded by the absence of chemical
potential pinning. We note that a recent Monte Carlo simulation has
shown that the introduction of disorder (by the alkaline-earth atom
substitution) replaces the chemical potential pinning by a smooth
shift. In that case, cluster formation on nanometer-scale occurs while
large cluster formation is still inhibited~\cite{Moreo2}.

Then let us further analyze the chemical potential shift in the
metallic regime $x\geq0.18$. For a metallic system, the rigid-band
picture predicts that
\begin{equation}
  \label{eq:1}
  \frac{\partial \mu}{\partial n}=\frac{1+F}{N^*({\mu})},
\end{equation}
where $N^*(\mu)=(3/\pi^2)(\gamma/k_B^2)$ is the density of states
(DOS) of renormalized quasiparticles (QP's) and $F$ is a parameter
which represents the effective QP-QP repulsion. Since
$\gamma=(\pi^2/3)k^2_BN^*(\mu)$, one can predict $\Delta\mu$ from the
measured $\gamma$, if $F$ is known.  $\Delta\mu$ predicted for $F=0$
is shown by the dotted curve in Fig.~\ref{fig:specific}. One can see
fairly good correspondence between the shift predicted by $\gamma$ and
the shift obtained by the XPS for $0.2 \leq x\leq0.4$. If we use
$F\simeq 1$, agreement between the calculated and measured shifts is
much improved. This means that the rigid-band description is
appropriate in the metallic phase. The specific heat
measurement~\cite{Okuda} indicated that the effective mass $m^*$ of
the quasiparticle is only slightly enhanced toward the metal-insulator
boundary and our result supports this observation.  Below $x \sim$
0.15, the $\Delta\mu$ calculated from $\gamma$ shows a rapid upturn
because $\gamma$ decreases in the insulating phase. In the insulating
phase ($x\leq0.15$), a rapid shift is expected if a band gap of finite
magnitude opens or the DOS at $\mu$ decreases.  However, the observed
shift $\partial\mu/\partial n$ is almost the same as in the metallic
region. This suggests that the insulating ground state has a finite
DOS at $\mu$ which are probably Anderson-localized.  

In conclusion, we have studied the chemical potential shift of
La$_{1-x}$Sr$_x$MnO$_3$ by means of x-ray core-level photoemission
spectroscopy. The shift is monotonous with hole concentration, which
suggests that there is no electronic phase separation on a macroscopic
scale that has been suggested by the electron microscopy
studies~\cite{Uehara}. On the other hand, the present observation is
consistent with the formation of nano-meter scale clusters induced by
chemical disorder. Comparison between the observed shift and the shift
predicted by the specific heat indicates that the hole doping in
La$_{1-x}$Sr$_x$MnO$_3$ is well described by the rigid-band picture.
Also, the mass enhancement toward the metal-insulator boundary was not
clearly observed through the chemical potential shift measurements,
qualitatively consistent with the specific heat data.

This work was supported by Grants-in-Aid for Scientific Research in
Priority Area ``Novel Quantum Phenomena in Transition-Metal Oxides"
and A12304018, and a Special Co-ordination Fund for the Promotion of
Science and Technology from the Ministry of Education, Culture,
Sports, Science and Technology of Japan.

\end{document}